\documentclass[usenatbib]{mn2e}

\usepackage{amsmath}
\usepackage{graphicx}
\usepackage{deluxetable}
\usepackage{savesym}
\usepackage{amsmath}
\savesymbol{iint}
\usepackage{txfonts}

\def\simless{\mathbin{\lower 3pt\hbox
{$\rlap{\raise 5pt\hbox{$\char'074$}}\mathchar"7218$}}}   
\def\simmore{\mathbin{\lower 3pt\hbox
{$\rlap{\raise 5pt\hbox{$\char'076$}}\mathchar"7218$}}}   
\newcommand{\be}{\begin{equation}}
\newcommand{\ee}{\end{equation}}
\newcommand       \bea          {\begin{eqnarray}}
\newcommand       \eea          {\end{eqnarray}}

\def\simlt{\mathrel{\hbox{\rlap{\hbox{\lower4pt\hbox{$\sim$}}}\hbox{$<$}}}}
\def\simgt{\mathrel{\hbox{\rlap{\hbox{\lower4pt\hbox{$\sim$}}}\hbox{$>$}}}}
\def\lesssim{\mathrel{\hbox{\rlap{\hbox{\lower4pt\hbox{$\sim$}}}\hbox{$<$}}}}

\def\gtrsim{\mathrel{\hbox{\rlap{\hbox{\lower4pt\hbox{$\sim$}}}\hbox{$>$}}}}

\title[Shock-powered emission in stellar mergers]{Shock-powered light curves of luminous red novae as signatures of pre-dynamical mass loss in stellar mergers}

\author[Metzger \& Pejcha]{Brian D.~Metzger$^{1}$, Ond\v{r}ej Pejcha$^{2}$\\
 $^{1}$ Columbia Astrophysics Laboratory, Columbia University, New York, NY, 10027, USA\\
 $^{2}$ Lyman Spitzer Jr.\ Fellow, Department of Astrophysical Sciences, Princeton University, Princeton, NJ, 08544, USA}

\begin{document}
\date{Received / Accepted}
\pagerange{\pageref{firstpage}--\pageref{lastpage}} \pubyear{2017}

\maketitle

\label{firstpage}

\begin{abstract}
Luminous red novae (LRN) are a class of optical transients believed to originate from the mergers of binary stars, or ``common envelope'' events.  Their light curves often show secondary maxima, which cannot be explained in the previous models of thermal energy diffusion or hydrogen recombination without invoking multiple independent shell ejections. We propose that double-peaked light curves are a natural consequence of a collision between dynamically-ejected fast shell and pre-existing equatorially-focused material, which was shed from the binary over many orbits preceding the dynamical event. The fast shell expands freely in the polar directions, powering the initial optical peak through cooling envelope emission. Radiative shocks from the collision in the equatorial plane power the secondary light curve peak on the radiative diffusion timescale of the deeper layers, similar to luminous Type IIn supernovae and some classical novae.  Using a detailed 1D analytic model, informed by complementary 3D hydrodynamical simulations, we show that shock-powered emission can explain the observed range of peak timescales and luminosities of the secondary peaks in LRN for realistic variations in the binary parameters and fraction of the binary mass ejected. The dense shell created by the radiative shocks in the equatorial plane provides an ideal location for dust nucleation consistent with the the inferred aspherical geometry of dust in LRN.  For giant stars, the ejecta forms dust when the shock-powered luminosity is still high, which could explain the infrared transients recently discovered by {\it Spitzer}.  Our results suggest that pre-dynamical mass loss is common if not ubiquitous in stellar mergers, providing insight into the instabilities responsible for driving the binary merger. 

\end{abstract} 
  

\section{Introduction}
Direct interactions between the component stars in a binary is a common process (e.g.~\citealt{Sana+12}) with key implications for all stages of stellar evolution, ranging from the proto-stellar phase (e.g.~\citealt{Krumholz&Thompson07}) to the death of low mass stars in proto-planetary nebulae (e.g., \citealt{Morris90}) and high mass stars in supernovae and gamma-ray bursts (e.g., \citealt{Cantiello+07,deMink+13}).  When at least one star begins overflowing its Roche lobe, the ensuing mass transfer may be unstable, resulting in a runaway process that culminates in both stars begin engulfed in a non-corotating ``common envelope" of gas (\citealt{Paczynski76,Webbink84,Livio&Soker88,Podsiadlowski01,Ivanova+13b}).  The merger process involves the ejection of mass and angular momentum from the system, leaving the final binary in a more compact configuration than its original separation or resulting in complete coalescence into a single star.  Though brief, this evolutionary stage plays an important role in the production of mass-transferring binaries, stripped-envelope or Type Ia supernova progenitors, and gravitational wave sources (e.g.~\citealt{Podsiadlowski+02,Postnov&Yungelson14}).  

The common envelope phase of stellar mergers has been studied using multi-dimensional hydrodynamical simulations of the dynamical inspiral stage (\citealt{Rasio&Livio96,Sandquist+88,Passy+12,Ricker&Taam12,Nandez+14,Nandez+15,Ohlmann+16,Ivanova&Nandez16})
and one-dimensional models which aim to follow the more slowly evolving (many dynamical time) phases of the process (\citealt{Taam+78,Meyer&MeyerHofmeister79,Ivanova02,Hall15}).  However, despite this extensive effort, our theoretical understanding of stellar mergers remains woefully incomplete, with even the most fundamental questions still unanswered (\citealt{Ivanova+13b}). 

One of these open issues relates to the initial cause of the instability, and how it connects to the final merger outcome.  Even prior to the onset of Roche lobe overflow, the binary may be driven together by secular tidal instabilities (e.g.~\citealt{Rasio&Shapiro92,Lai+93}), such as the Darwin instability (e.g.~\citealt{Rasio95}).  However, this process is generally slow and will likely be overtaken by other processes once mass transfer begins.  If one star fills its Roche lobe before the other, dynamically unstable mass transfer can occur, depending on the non-adiabatic response of the surface layers of the mass-transferring star (e.g.~\citealt{Hjellming&Webbink87,Ge+10,Passy+12}), filling the common Roche surface with substantial gaseous mass.  If the envelope rotates slower than the binary, the resulting hydrodynamical drag of the binary orbiting in this ``egg beater'' configuration provides one mechanism to extract angular momentum and drive the stars together (e.g.~\citealt{Macleod+17b}b).  However, given the low inertia of the envelope, co-rotation with the orbital velocity could well be maintained during the earliest stages of the merger.  

An alternative mechanism for driving the binary together is the loss of mass and angular momentum through the outer Lagrange points, especially L2 (\citealt{Flanner&Ulrich77,Shu+79,Lombardi+11,Pejcha14,Pejcha+16a,Pejcha+16b}).  Regardless of whether mass loss provides a dominant sink of angular momentum driving the merger, it may nevertheless accompany the process prior to the dynamical inspiral.  Even beginning ``cold", mass near the L2 point can be unbound by non-axisymmetric torques from the binary \citep{Shu+79}.  This produces a sprinkler-like outflow focused in the orbital plane, with a mass loss rate that increases in strength approaching the final coalescence event \citep{Pejcha+16a,Pejcha+16b}. 
If observable, signatures of this pre-dynamical mass loss could be diagnostic of the instabilities responsible for driving the binary together and the mass loss history prior to the final dynamical stage \citep{Pejcha14}.  It is clearly important to explore what observational constraints exist on the presence of pre-dynamical mass loss in stellar mergers.

\begin{figure*}
  \centering
  \includegraphics[width=0.8\textwidth]{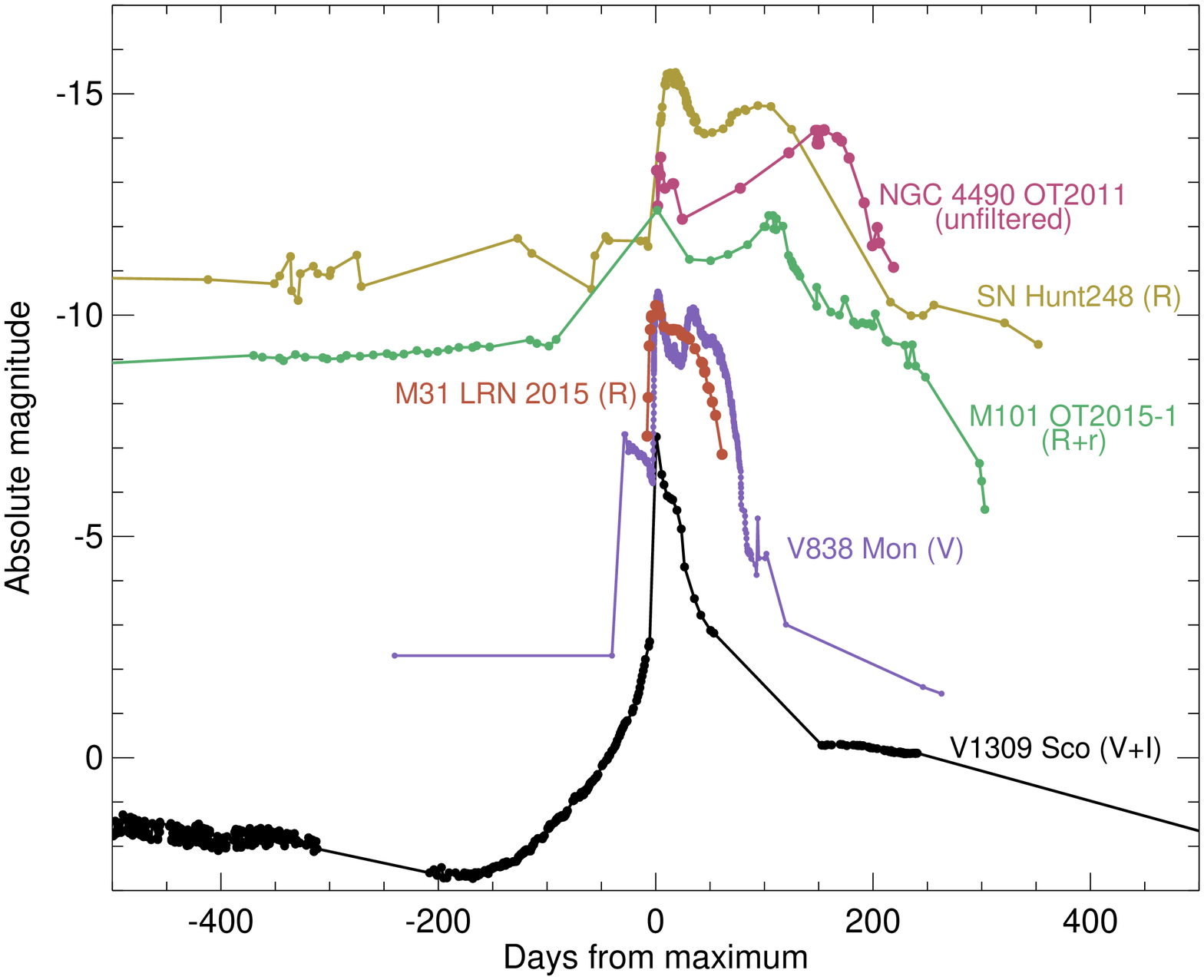}
  \caption{Light curves of stellar merger candidates.  Events shown include NGC 4490 OT2011 (green; \citealt{Smith+16}), SN Hunt248 (pink; \citealt{Mauerhan+15}, \citealt{Kankare+15}), M101 OT2015-1 (brown; \citealt{Blagorodnova+17}), V838 Mon (yellow; AAVSO International Database), V1309 Sco (black; \citealt{Tylenda+11}, \citealt{Pojmanski02}, AAVSO International Database), M31 LRN 2015 (green; \citealt{Kurtenkov+15}). We show single-band light curves corrected for reddening. Wiggles seen in some objects are due to uncorrected differences between different data sources.}
 \label{fig:LC}
\end{figure*}

An emerging class of optical transients, commonly known\footnote{Alternative names used in the literature include `intermediate luminosity optical transients' and `mergebursts' \citep{Soker&Tylenda06}.} as ``luminous red novae" (LRN; \citealt{Martini+99,Munari+02,Bond+03,Mauerhan+15,Kankare+15,Kurtenkov+15,Smith+16,Blagorodnova+17}), are believed to originate from stellar merger events \citep{Soker&Tylenda03,Tylenda+05,Soker&Tylenda06,Tylenda&Soker06,Smith11,Ivanova+13a,Nandez+14,Macleod+17}.  LRN are characterized by luminosities intermediate between those of classical novae and supernovae, and by their comparatively red colors at late times (which also distinguish them from classical novae, which evolve to the blue).  They are estimated to occur every $\sim 2$ years in Milky-Way type galaxies \citep{Kochanek+14}.  A sample of LRN light curves is shown in Figure~\ref{fig:LC}.  Despite significant diversity in their duration and peak luminosity, they are generally characterized by a slow but accelerating rise to the a first peak, followed by either a plateau-like stage or a more gradual rise to a second maxima.  The true duration of the outburst probably lasts longer than indicated by the optical light curve, due to the inferred formation of dust, and shift of the SED to the infrared, at late times.      

The watershed event confirming the association of LRN with stellar mergers was the galactic transient V1309 Sco \citep{Mason+10,Tylenda+11}.  A multi-year time series of photometric data from OGLE revealed an eclipsing binary with a decreasing orbital period prior to the outburst \citep{Tylenda+11,Stepien11}.  Following the event, the optical variability disappeared \citep{Tylenda+11,Kaminski+15}, but the object remains enshrouded in dust \citep{Nicholls+13}.  \citet{Pejcha14} interpreted the pre-explosion brightening of this event as the result of L2 mass loss, with the emission being powered primarily by heating due to internal shocks between the L2 spirals accompanied by dust production \citep{Pejcha+16a,Pejcha+16b}.  Indeed, the pre-explosion SED shows evidence for dusty mass loss already during the final inspiral phase \citep{Tylenda&Kaminski16}.  This pre-explosion brightening is observed in other LRN, such as M101 OT2015-1 \citep{Blagorodnova+17} and M31 LRN 2015 (\citealt{Dong+15,Williams+15}), suggesting that pre-dynamical mass loss may be ubiquitous.

\begin{figure}
  \centering
  \includegraphics[width=1.0\columnwidth]{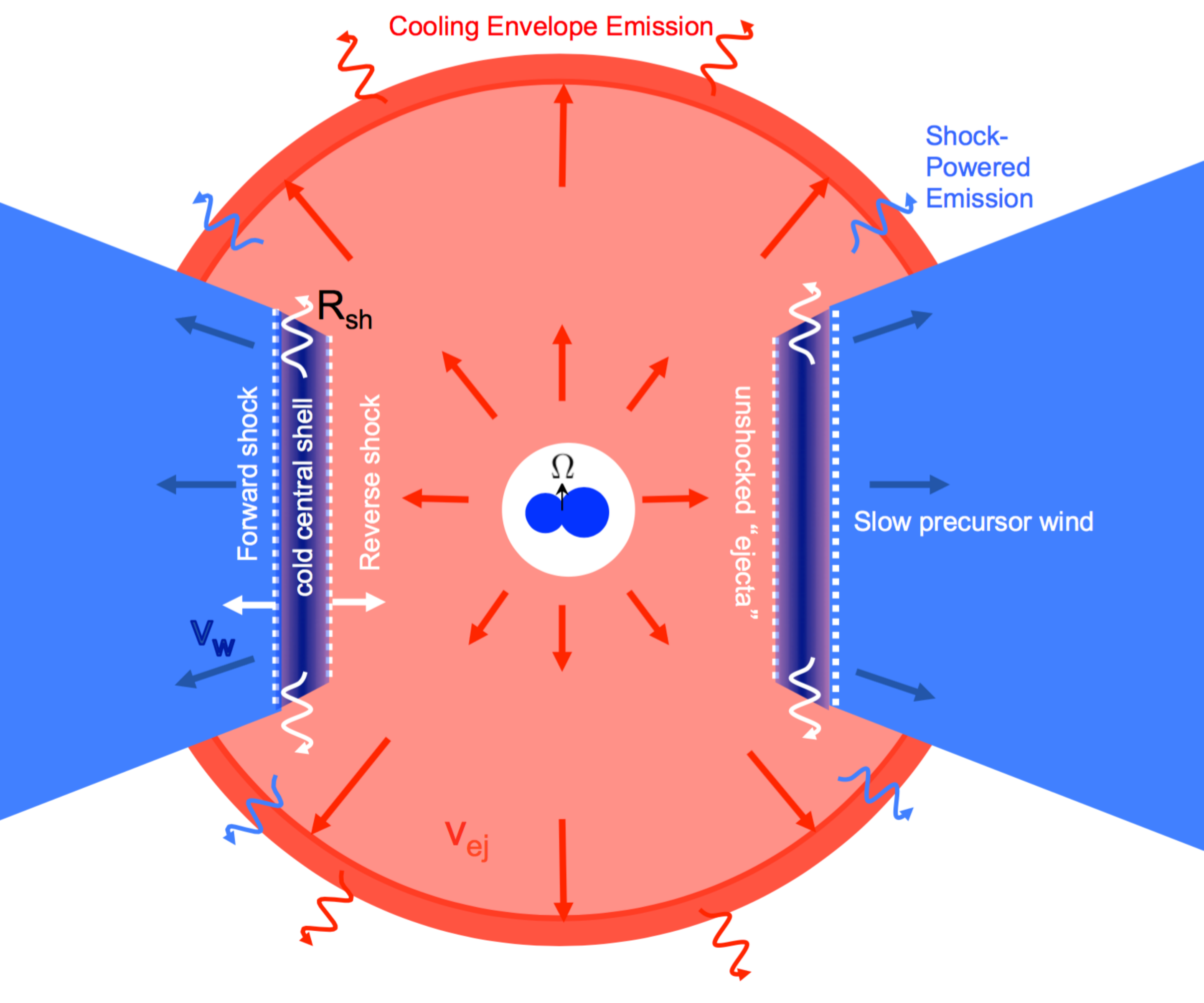}
  \caption{Schematic illustration of shocks in stellar mergers and the power source of LRN.  During the terminal phase of the merger, a fast dynamically-ejected shell (red) collides with with a slow equatorially-focused wind (blue) from the pre-dynamical L2 mass loss phase.  The fast ejecta expands freely in the polar direction, releasing the initial thermal energy of the explosion in cooling envelope emission and powering the first peak in the LRN light curve.  Radiative shocks between the fast and slow outflows in the equatorial plane produce radiation, which first diffuses out through the disk in the vertical direction and then radially through the fast ejecta, powering the second light curve peak.  Gas swept up by the shocks collects in a cool, dense, shell, providing a natural location for subsequent dust formation in the ejecta.  Dust formation may occur earlier in the merger of giant stars, allowing a sizable fraction of the shock power to be reprocessed by the dust itself and producing a luminous infrared transient.}
 \label{fig:cartoon}
\end{figure}


\begin{table*}
\caption{Properties of LRN and their stellar progenitors \label{table:1}}
\begin{tabular}{lccccccc}
\hline
\hline
{Event}                   &
{$M$}  &
{$R_{\star}$}           &
{$v_{\rm esc}$}    &
{$v_{\rm ej}^{(a)}$} & 
{$L_{\rm pk}^{(b)}$} &
{$t_{\rm pk}^{(b)}$} & 
{Refs} \\
{} &
{($M_{\odot}$)}  &
{($R_{\odot}$)}      &
{(km s$^{-1}$)}       &
{(km s$^{-1}$)} & 
{(erg s$^{-1}$)} &
{(days)} &
{}\\
\hline
M31 LRN 2015 & 3$-$5.5 & 30$-$40 & $100-130$ & 370$\pm 240$ & N/A & N/A & 8\\
SN Hunt248 & 30$\pm 2$ & 500 & 75 & 1200 & $6\times 10^{40}$ $^{(c)}$ & 100 & 6\\
V838 Mon & 5$-10$ & 4-6 & 350$-$400 & 500  & $4\times 10^{39}$ & 40 & 2, 3 \\
V1309 Sco & 1.5 & 3.5 & 170$-$500 & 150 & N/A & N/A & 1 \\
M101 OT2015-1  & 18  & 220 & 180 & 500 & $3\times 10^{40}$ & 110 & 4 \\
NGC 4490-OT & 30 & - & 280-650 & - & $7 \times 10^{40}$ & 160 & 7 \\
\hline
\end{tabular}
\\
{$^{(a)}$Ejecta velocity inferred from optical spectra (note that this may not always represent the velocity of the bulk of the ejecta, but instead the fastest components).  $^{(b)}$Approximate peak luminosity of second peak.}
$^{(c)}$Assuming the second peak was bolometrically about 1 mag fainter than the first\\
References:
  (1) \citet{Mason+10}, (2) \citet{Munari+02}, (3) \citet{Tylenda+05,Afsar&Bond07} but see \cite{Munari+05}, (4) \citet{Blagorodnova+17}, (5) \citet{Smith+16} , (6) \citet{Mauerhan+15}, (7) \citet{Mauerhan+17}, (8) \citet{Macleod+17}
\end{table*}

The optical emission from LRN was previously proposed to originate from hydrogen recombination in a shell, or multiple shells, of mass ejected during the dynamical stage of the merger process \citep{Ivanova+13b,Lipunov17}, similar to the physics controlling the plateau phase of Type IIP supernova light curves.  However, \citet{Macleod+17} found that recombination energy cannot simultaneously explain the short duration and high luminosities of the first light curve peak of M31 LRN 2015 given the characteristic velocities observed.  They argue that the first peak could instead be explained by the release of thermal emission from the hot ejecta, similar to the ``cooling envelope" phase of early supernova emission (e.g., \citealt{Nakar&Sari10}).  On the other hand, a single energy source cannot readily explain the doubled-peaked shape of many LRN light curves (Fig.~\ref{fig:LC}), without invoking multiple distinct shell ejection events separated by time intervals of many binary orbital periods.  Also mysterious is the complex color evolution in some LRN (e.g., SN Hunt248, which evolves from red to blue and then back to red, Fig.~3 of \citealt{Mauerhan+15}; M31 LRN 2015, Figs.~1 and 2 of \citealt{Kurtenkov+15}), which contrasts with the monotonic blue to red evolution of most supernovae.

Here we propose a new model for the optical emission of LRN, based on the shock interaction between the dynamically-ejected mass and a prior phase of equatorially-concentrated mass loss.  Figure \ref{fig:cartoon} provides a schematic illustration of the idea.  Our proposed mechanism is qualitatively similar to that invoked to explain the light curves of luminous Type IIn supernovae, powered by the collision between the supernova ejecta and pre-explosion stellar mass loss (e.g., \citealt{Smith&McCray07,Chevalier&Irwin11}) and classical novae \citep{Metzger+14,Metzger+15}.  In the polar direction, where the density of the pre-dynamical medium is lowest, the fast dynamical ejecta will expand freely.  The first light curve maximum is powered via diffusion of thermal emission from these hot outermost layers (``cooling envelope"; \citealt{Macleod+17}).  In the equatorial plane of the binary the fast ejecta collides with the slow equatorial wind; these shocks provide a sustained source of radiation which powers the second light curve peak (the light curve peaks later because the shocks are deeply embedded and hence the radiative diffusion timescale is longer).  The combination of cooling envelope and shock-powered emission thus provides a natural explanation for the double-peaked light curves of many LRN.  

Our model shows that the secondary maxima of LRN light curves do not require an additional dynamical mass ejection events, but instead imprint valuable information on the mass loss history prior to the dynamical merger phase.  The common presence of luminous secondary peaks (Fig.~\ref{fig:LC}) thus indicates that the growth rate of pre-dynamical mass loss is generally gradual, which in turn suggests that the instability responsible for the merger initiated many orbital periods prior to the final dynamical phase (\citealt{Pejcha14}).  

Beyond its implications for LRN optical emission, our model suggests a natural geometry for dust formation in stellar merger ejecta as being concentrated in the equatorial plane where the gas is swept into a dense shell by highly compressible radiative shocks, as supported by observations \citep{Kaminski+10,Chesneau+14,Mauerhan+17}.  For particularly wide binary mergers, we find that dust may form at an early stage, in which case the shock-powered emission could be completely embedded within, and reprocessed by, dust.  This provides a possible explanation for the class of infrared bright (but optically obscured) transients recently discovered as part of the recent {\it Spitzer} survey of nearby galaxies \citep{Kasliwal+17}.  Particularly early dust formation may occur in mergers where the shocks are insufficient to keep the ejecta ionized, possibly explaining the absence of clear secondary optical light curve peaks in V1309 Sco \citep{Tylenda+11} and M31 LRN 2015 \citep{Kurtenkov+15}.

This paper is organized as follows.  In $\S\ref{sec:numerical}$ we motivate the basic physical picture using SPH hydrodynamical simulations of the interaction between a fast dynamically-ejected shell and a slow  equatorially-concentrated wind.  In $\S\ref{sec:analytic}$ we describe a 1D numerical model for the shock interaction and a simple radiative diffusion model for LRN light curves.  In $\S\ref{sec:results}$ we describe results of the 1D model, present some analytic scalings ($\S\ref{sec:scaling}$), and describe the implications for dust formation in stellar mergers ($\S\ref{sec:dust}$).  In $\S\ref{sec:discussion}$ we discuss our results and summarize our conclusions.

\section{Mass Loss: Slow and Fast Outflows}
\label{sec:numerical}

\subsection{Initial conditions}
\label{sec:initial}

\begin{figure*}
  \includegraphics[width=2.0\columnwidth]{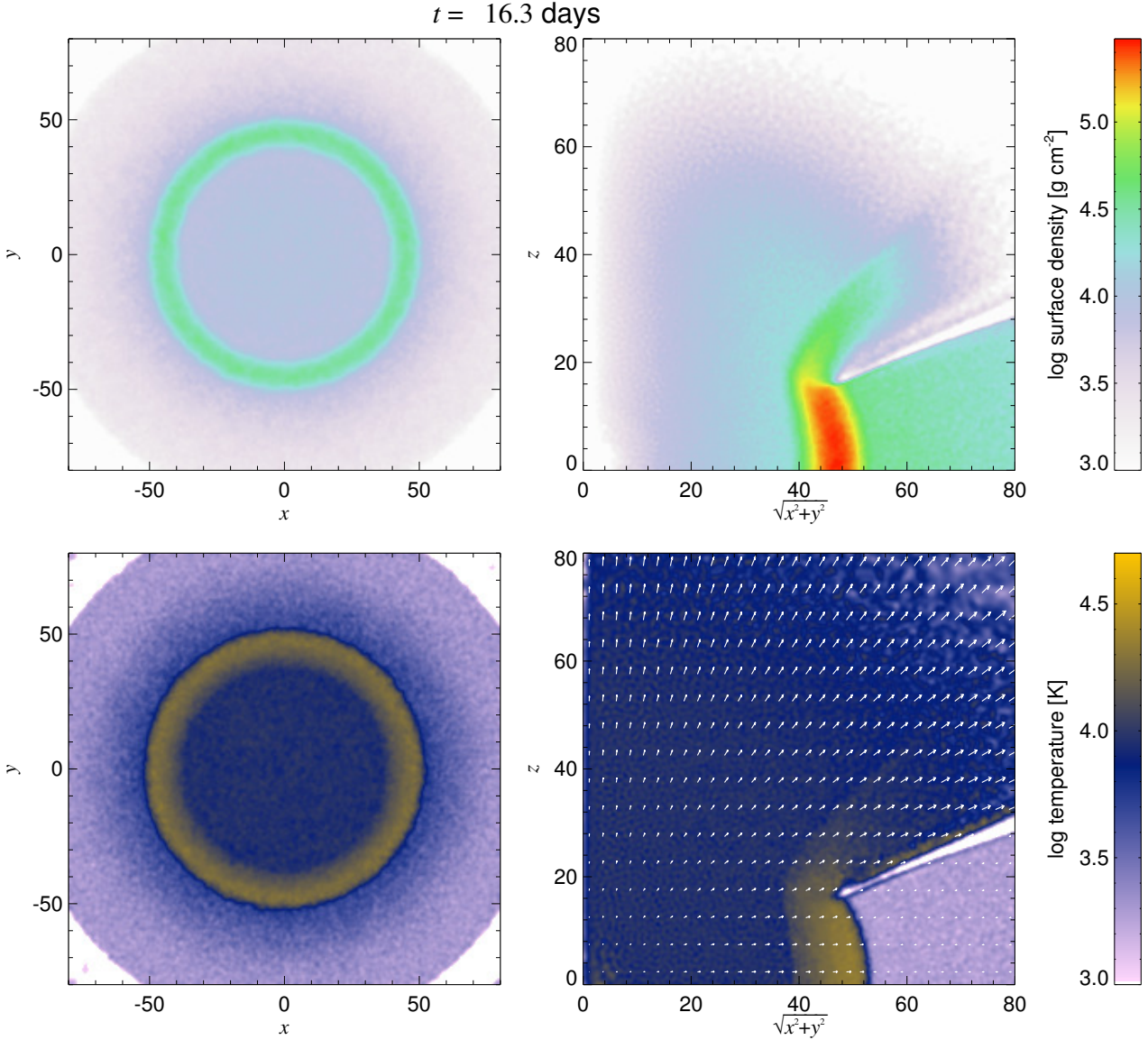}
  \caption{Density and temperature structure of the collision of initially spherical homologuously expanding ejecta and pre-existing equatorially-concentrated outflow. The two left panels show projections of density and temperature in the binary orbital plane, while the two right panels show the vertical structure of the same obtained by projecting the particles in the $\sqrt{x^2+y^2}-z$ plane. The arrows in the lower right panel denote the velocity vectors. We assumed a system with $a=0.03$\,AU, $M=1.65\,M_\odot$, $M_{\rm ej} = M_{\rm w} = 0.01\,M_\odot$ distributed among a total of $2\times 10^6$ equal-mass particles. The equatorial outflow had $v_w = 50$\,km\,s$^{-1}$, half-opening angle of 20 degrees, $t_{\rm run} = 160 t_{\rm orb}$, and $T=2000$\,K. The central hot ball was initiated with $T = 5 \times 10^5$\,K. The animated version of this figure is available in the on-line version or at \tt{https://youtu.be/3IG9pJhRFR8}.}
 \label{fig:numerical}
\end{figure*}

Consider the merger of a stellar binary of semi-major axis $a$, total mass $M$, mass ratio $q \le 1$, and escape speed $v_{\rm esc} \approx (2GM/a)^{1/2}$.  Prior to the final, dynamical stage of the event, an external density profile is laid down by the prolonged prior phase of equatorially-focused mass loss.  We approximate this outflow as that due to a wind of constant velocity $v_{\rm w} \lesssim v_{\rm esc}$, 
\be
\rho_{\rm w} = \frac{\dot{M}(t_{\rm w} = r/v_{\rm w})}{4\pi f_{\Omega} v_{\rm w}r^{2}},
\label{eq:rhow}
\ee
where $\dot{M}$ is the wind mass-loss rate at time $t_{\rm w}$ prior to the dynamical explosion and $f_{\Omega} \approx h/r \approx 0.1-0.3$ is the fraction of the solid angle in the equatorial plane subtended by the wind, where $h$ is the vertical height of the slow ejecta.  We parametrize the pre-dynamical mass loss rate through a simple prescription
\be
\dot{M}(t_{\rm w}) = \frac{M_{\rm w}}{t_{\rm run}} e^{-t_{\rm w}/t_{\rm run}},
\label{eq:Mdot}
\ee 
where $M_{\rm w}$ is the total wind mass loss, $t_{\rm run} = Nt_{\rm orb}$ and $N$ is the number of binary orbital periods $t_{\rm orb} = 2\pi(a^{3}/GM)^{1/2}$ over which the mass loss grows approaching the merger. For V1309~Sco one infers $N \sim 5$ to $20$ immediately prior to the outburst \citep{Tylenda+11,Pejcha14}, though the evolution might be more complex than captured by Equation~(\ref{eq:Mdot}).  If the pre-dynamical mass loss leaves L2 with roughly the specific angular momentum of corotation, we can obtain an estimate of $M_{\rm w}$. The long lever arm of L2 implies that a binary needs to lose only $10$ to $15\%$ of the secondary mass $qM/(1+q)$ to get rid of all of its orbital angular momentum \citep{Macleod+17}. The required mass loss is obviously greater if the specific angular momentum is lower than that of L2.

We assume that the dynamical phase of the merger is characterized by the (likely shock-driven) ejection of a total mass  $M_{\rm ej}$ released effectively instantaneously at $t = 0$ from a radius $r_{\rm in} \approx a$.  The dynamical ejecta is spherically symmetric with homologous radial profile, for which the velocity at radius $r$ is given by $v_{\rm ej} = r/t$.  We model the density distribution as that of a relatively flat inner profile and a steeper power-law tail of high velocity ejecta 
\begin{eqnarray}
\rho_{\rm ej} = 
\left\{
\begin{array}{lr}
\frac{(\delta_{\rm o} - 3)(3-\delta_{\rm i})}{4\pi(\delta_{\rm o} - \delta_{\rm i})}\frac{M_{\rm ej}}{(\bar{v}_{\rm ej}t)^{3}}\left(\frac{v_{\rm ej}}{\bar{v}_{\rm ej}}\right)^{-\delta_{\rm i}},
 & r \le \bar{v}_{\rm ej}t
\\
\frac{(3-\delta_{\rm i})(\delta_{\rm o}-3)}{4\pi(\delta_{\rm o} - \delta_{\rm i})}\frac{M_{\rm ej}}{(\bar{v}_{\rm ej}t)^{3}}\left(\frac{v_{\rm ej}}{\bar{v}_{\rm ej}}\right)^{-\delta_{\rm o}}, &
r > \bar{v}_{\rm ej}t. \\
\end{array}
\right.
\label{eq:rhoej}
\end{eqnarray}
where $\bar{v}_{\rm ej} \approx v_{\rm esc}$ approximately equals the mean ejecta velocity and the power-law indices of the inner ejecta ($\delta_{\rm i} < 3$) and outer ejecta ($\delta_{\rm o} > 3$) depend on the details of the dynamical phase (\citealt{Chevalier&Soker89}).  The mass above a velocity $v_{\rm ej}$ is given by
\begin{eqnarray}
M_{\rm v} = 
\left\{
\begin{array}{lr}
\frac{(3-\delta_{\rm i})M_{\rm ej}}{(\delta_{\rm o} - \delta_{\rm i})} + \frac{(\delta_{\rm o} - 3)M_{\rm ej}}{(\delta_{\rm o} - \delta_{\rm i})}\left[1-\left(\frac{v}{\bar{v}_{\rm ej}}\right)^{3-\delta_{\rm i}}\right],
 & r \le \bar{v}_{\rm ej}t
\\
\frac{(3-\delta_{\rm i})M_{\rm ej}}{(\delta_{\rm o} - \delta_{\rm i})}\left(\frac{v}{\bar{v}_{\rm ej}}\right)^{3-\delta_{\rm o}}. &
r > \bar{v}_{\rm ej}t. \\
\end{array}
\right.
\label{eq:Mv}
\end{eqnarray}
We have few a priori constrains on $M_{\rm ej}$ apart from the usual energy arguments applied to stellar mergers \citep[e.g.][]{Webbink84}. The actual value of $M_{\rm ej}$ must depend on the masses and structure of the binary components. To explain the observed light curves we typically require $M_{\rm ej} \sim M_{w}$.

\subsection{Hydrodynamics of the interaction}
\label{sec:hydro}

To motivate the analytic model developed in Section~\ref{sec:analytic}, we perform several hydrodynamical simulations of this problem. We use a simplified version of the smoothed particle hydrodynamics (SPH) code developed in \citet{Pejcha+16a,Pejcha+16b}, which includes standard hydrodynamical and viscous evolution \citep{Price+07,Monaghan+83,Balsara95} with a realistic equation of state \citep{tomida13}. We neglect gravitational forces and radiative processes. We set up the initial conditions by distributing cold equal-mass particles in the equatorial outflow according to the density distribution of Equation~(\ref{eq:rhow}) with an inner boundary at $5a$ and outer boundary at $80a$. The dynamical ejecta is initiated by placing particles in rest within a sphere of radius $2a$ and setting their temperature to a high value. The expansion of the central hot ball converts its internal energy to kinetic energy and naturally sets up homologous expanding velocity profile, although we do not attempt to match Equation~(\ref{eq:rhoej}). Although initiating the simulation with particles on a regular grid leads to better energy conservation, for the visualization purposes we show qualitatively identical results from a run with randomly positioned particles.

Figure~\ref{fig:numerical} shows results from one of the simulations. The dynamical ejecta is free to expand in the polar direction, but near the equatorial plane, where the expansion slows down due to the pre-existing material, a torus of swept-up material bounded by the forward and reverse shocks quickly develops.  As the torus plows through the equatorial outflow, a fraction of the swept-up material is squeezed above and below into the homologously expanding ejecta, where it is accelerated to higher velocities. This leads to an overdensity of material at mid-latitudes, which might be the cause of ring-like features seen in some merger remnant candidates such as the progenitor of SN1987A \citep{Morris07}. 

In this highly idealized setting, the two components interact only near the torus of swept-up material, as envisioned in Figure~\ref{fig:cartoon}. We experimented with more realistic settings appropriate for V1309~Sco and found that the morphology of the interaction region depends on the details of the density profile. Specifically, the tilt of the forward/reverse shock with respect to the orbital axis can vary depending on the vertical density structure of the equatorial outflow. In some cases we saw finger-like instabilities in the swept-up torus, like in other similar settings \citep[e.g.][]{Chen16}. However, here we focus on the implications of this general structure for the light curves and leave the detailed investigation of the hydrodynamics to future work.

\section{Analytic Model}
\label{sec:analytic}

\subsection{Shock Interaction}

After the onset of the explosion ($t > 0$), the fast ejecta drives a forward shock into the slow pre-dynamical wind, while a reverse shock passes through the fast ejecta (Figs.~\ref{fig:cartoon}, \ref{fig:numerical}).  Given the high gas densities and characteristic shock velocities of hundreds of km s$^{-1}$ or less, the shocks are radiative (cooling time much shorter than expansion timescale) and the post-shock gas collects into a thin dense shell of radius $R_{\rm sh}(t)$.  The mass of the cold shell grows as the result of sweeping up both winds by radiative shocks (\citealt{Metzger+14})
\be
\frac{dM_{\rm sh}}{dt} = 4\pi f_{\Omega}R_{\rm sh}^{2}\rho_{\rm ej,sh}\left(\frac{R_{\rm sh}}{t}-v_{\rm sh}\right) + 4\pi R_{\rm sh}^{2}\rho_{\rm w,sh}(v_{\rm sh}-v_{\rm w})
\label{eq:dMdt}
\ee
while the swept up momentum of the shell grows as
\be
\frac{d}{dt}\left(M_{\rm sh}v_{\rm sh}\right) =  4\pi f_{\Omega}R_{\rm sh}^{2}\rho_{\rm ej,sh}\frac{R_{\rm sh}}{t}\left(\frac{R_{\rm sh}}{t}-v_{\rm sh}\right) + 4\pi R_{\rm sh}^{2}\rho_{\rm w,sh}v_{\rm w}(v_{\rm sh}-v_{\rm w}),
\label{eq:dPdt}
\ee
where $\rho_{\rm ej,sh}(t) \equiv \rho_{\rm ej}(R_{\rm sh})$ and $\rho_{\rm w,sh}(t) \equiv \rho_{\rm w}(R_{\rm sh})$ are the density of the slow wind and the fast ejecta just upstream of the shocks.

The power dissipated at the forward and reverse shocks from the jump conditions are given, respectively, by
\be
L_{\rm f} = \frac{9\pi}{8}R_{\rm sh}^{2}\rho_{\rm w,sh}(v_{\rm sh}-v_{\rm w})^{3}
\label{eq:Lf}
\ee
\be
L_{\rm r}  = \frac{9\pi }{8}f_{\Omega}R_{\rm sh}^{2}\rho_{\rm ej,sh}\left(\frac{R_{\rm sh}}{t} - v_{\rm sh}\right)^{3}
\label{eq:Lr}
\ee
This power is primarily released behind the shock as UV line emission.  However, given the high columns of gas ahead of the shocks and in the central shell, this hard radiation will be absorbed and reprocessed to lower, optical frequencies.  

\subsection{Light Curve Model}

For optical radiation from the shocks to escape to an external observer, it must diffuse both out through the slow equatorial outflow into the polar region occupied by the fast ejecta, and then radially through the fast ejecta (the initial thermal energy of the dynamical ejecta obviously passes just through the latter).  In order to calculate the light curve, we follow the evolution of the internal energy of the slow wind, $E_{\rm w}$, and that of the dynamical ejecta, $E_{\rm ej}$.  

We model the slow wind as a single zone near the shock of volume $V_{\rm w} = 4\pi R_{\rm sh}^{2}h$ and mass $V_{\rm w}\rho_{\rm sh,w}$, in which the specific internal energy $E_{\rm w}$ evolves according to
\be
\frac{dE_{\rm w}}{dt} = -\eta_{\rm w} \frac{E_{\rm w}}{R_{\rm sh}}\frac{dR_{\rm sh}}{dt} + \frac{(L_{\rm f} + L_{\rm r})}{V_{\rm w}\rho_{\rm w,sh}} - \frac{E_{\rm w}}{t_{\rm d,w}}.
\label{eq:Ew}
\ee 
The first term accounts for PdV losses, where the ratio $\eta_{\rm w} \equiv 3P_{\rm w}/E_{\rm w}$ varies between $1-2$ depending on whether radiation or gas dominates the pressure $P_{\rm w}$.  The source term is the luminosities of the shocks (eqs.~\ref{eq:Lf},\ref{eq:Lr}).  

The last term in equation (\ref{eq:Ew}) accounts for vertical radiative diffusion into the polar region occupied by the fast ejecta.  This occurs on the radiative diffusion timescale 
\be
t_{\rm d,w} = (\tau_{\rm w}+1)\frac{h}{2c} = (\rho_{\rm w,sh}\kappa_{\rm w}R_{\rm sh}f_{\Omega}+1)\frac{ R_{\rm sh}f_{\Omega}}{2c},
\ee 
where $\tau_{\rm w} = \rho_{\rm w,sh}\kappa_{\rm w}h$ is the vertical optical depth through the disk of scale-height $h = f_{\Omega}R_{\rm sh}$ and the factor of 2 accounts for diffusion through the top and bottom halves of the disk.  The Rosseland mean opacity of the disk $\kappa_{\rm w}$ depends on the local density $\rho_{\rm w,sh}$ and the temperature $T_{\rm w}$ (see eq.~\ref{eq:kappa} below); the latter is determined from the density and internal energy $E_{\rm w}$ according to the equation of state.  We adopt a simple equation of state which includes gas and radiation pressure.  

We simultaneously follow the evolution of the specific energy of the fast ejecta.  We divide the fast ejecta into a large number of discrete shells of velocity $v_{\rm ej}$ and mass $\Delta M_{\rm v} = M_{\rm v+\Delta v}-M_{\rm v}$ according to the density distribution given in equation (\ref{eq:Mv}).  The specific internal energy of each shell $E_{\rm ej,v}$ evolves according to
\be
\frac{dE_{\rm ej,v}}{dt} = -\eta_{\rm ej,v} \frac{E_{\rm ej,v}}{t} + \frac{E_{\rm w}}{t_{\rm d,w}}\frac{V_{\rm w}\rho_{\rm w,sh}}{\Delta M_{v}} - \frac{E_{\rm ej,v}}{t_{\rm d,v}},
\label{eq:Eej}
\ee 
where again the first term accounts for PdV losses where $\eta_{\rm ej,v} \equiv 3P_{\rm ej,v}/E_{\rm ej,v}$.  The second term accounts for the diffusion of shock energy into the fast ejecta, which we assume is distributed uniformly per unit mass throughout the fast ejecta.\footnote{In reality, the shocks will deposit most of their radiation locally near the mass shells corresponding to the instanteneous shock location.  However, our attempt to localize the heating led to unphysical light curve behavior at some epochs, due to our simple shellular treatment of the radiative transfer.  This simplification does not result in large quantitative errors for the light curve because the shock usually resides near the inner ejecta layers, the diffusion from which dominates the secondary light curve maximum. }  

The last term in equation (\ref{eq:Eej}) accounts for radiative diffusion in the radial direction through the ejecta, as occurs on the diffusion timescale
\begin{eqnarray}
t_{\rm d,v} = (\tau_{\rm ej,v}+1)\frac{R_{\rm ej}}{\delta \cdot c}
\end{eqnarray}
where $\tau_{\rm ej,v} = \rho_{\rm ej,v}\kappa_{\rm ej,v}R_{\rm ej}$ is the radial optical depth of the shell of radius $R_{\rm ej} = v_{\rm ej}t$, local density slope $\delta = \delta_{i}, \delta_{o}$ (eq.~\ref{eq:rhoej}), and Rosseland mean opacity $\kappa_{\rm ej, v}$.  The latter depends on the density $\rho_{\rm ej,v}$ and temperature $T_{\rm ej,v}$ of the shell, which is determined from $E_{\rm ej,v}$ based on the equation of state.   The diffusion time is generally a monotonically decreasing function of radius, but when this is not the case we artificially set the diffusion time to its maximum value above the given shell's velocity.

In Type IIP supernovae, the ejecta is relatively dilute and the opacity of fully ionized matter is well-approximated as that due to electron scattering.  However, the comparatively smaller velocities of stellar merger ejecta ($v \lesssim 500$\,km\,s$^{-1}$) imply much higher densities $\rho \propto M_{\rm ej}/v^3$, necessitating the use of a more complex opacity formulation.  We calculate the opacity of both the slow wind and fast ejecta using the following approximate analytic formula (for solar metallicity; $Z = 0.02, X = 0.74$)
\be
\kappa \approx \kappa_m + \left(\kappa_{\rm H^{-}}^{-1} + (\kappa_e + \kappa_{\rm K})^{-1}\right)^{-1} ,
\label{eq:kappa}
\ee  
which accounts for electron scattering $\kappa_e \approx 0.2(1+X)$ cm$^{2}$ g$^{-1}$, bound-free/free-free absorption $\kappa_{\rm K} \approx 4\times 10^{25}Z(1+X)\rho T^{-7/2}$ cm$^{2}$ g$^{-1}$, H$^{-}$ opacity $\kappa_{\rm H^{-}} \approx 1.1\times 10^{-25}Z^{0.5}\rho^{0.5}T^{7.7}$ cm$^{2}$ g$^{-1}$, and a characteristic molecular opacity $\kappa_m \approx 0.1Z$ cm$^{2}$ g$^{-1}$

Equations (\ref{eq:dMdt}, \ref{eq:dPdt}) are solved for the shock radius $R_{\rm sh}(t)$ and radiative luminosities $L_{\rm f}(t), L_{\rm r}(t)$.  These results then serve as input in solving the coupled equations (\ref{eq:Ew} and \ref{eq:Eej}).  As initial conditions, we assume that the wind material is cold and that the dynamical ejecta is shocked heated to a specific energy $E_{\rm ej,v}(t_0) = v_{\rm ej}^{2}/2$ at initial time $t_{0} = 0.1 t_{\rm dyn}$, where $t_{\rm dyn} = (a^{3}/GM)^{1/2}$.  The validity of this prescription is uncertain, and depends on the details of the dynamical merger phase and the shock-driven outflows.

The total bolometric luminosity is given by the sum of the luminosities of each mass layer
\be
L_{\rm bol}(t) = \sum_{\rm v}L_{\rm rad,v}(t) = \sum_{\rm v}\Delta M_{\rm v} \frac{E_{\rm ej,v}(t)}{t_{\rm d,v}(t)}
\ee
We approximate the photosphere radius by its luminosity-weighted value
\be
R_{\rm ph} = \frac{1}{L_{\rm bol}}\sum_{\rm v}L_{\rm rad,v} (v_{\rm ej}t)
\ee
The photosphere temperature is then given by
\be
T_{\rm ph} = \left(\frac{L_{\rm bol}}{4\pi \sigma R_{\rm ph}^{2}}\right)^{1/2}
\ee

\section{Results of Analytic Light Curve Model}
\label{sec:results}

\begin{figure}
  \centering
  \includegraphics[width=1.\columnwidth]{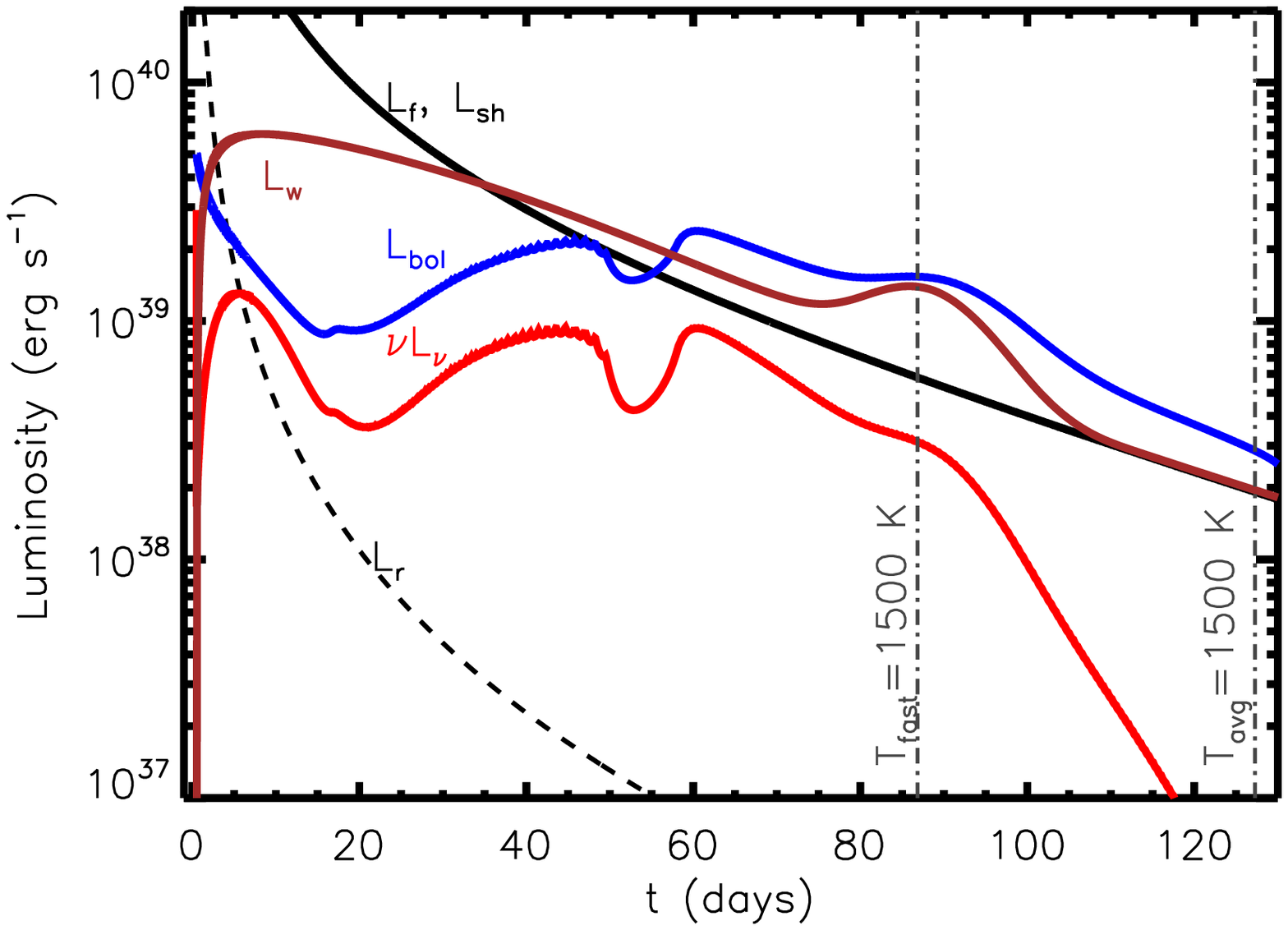}
  \includegraphics[width=1.\columnwidth]{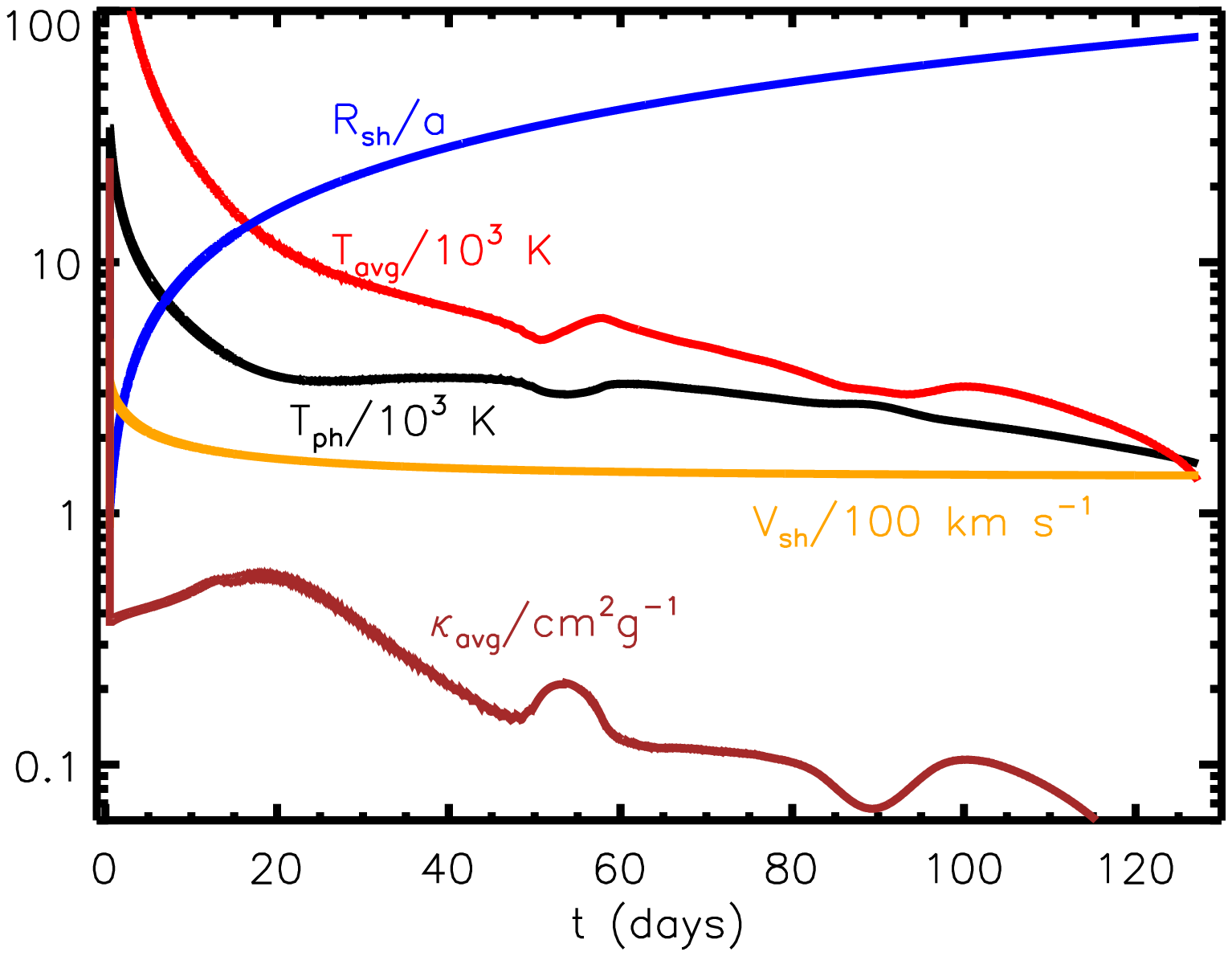}
  \caption{Example calculation for  $M = 10M_{\odot}$, $a = 31R_{\odot}$, $v_{\rm esc} = 350$ km s$^{-1}$, $v_{\rm w} = v_{\rm esc}/4$, $f_{\Omega} = 0.3$, $\bar{v}_{\rm ej} = v_{\rm esc}$, $\bar{v}_{\rm max} = 5v_{\rm esc}$, $M_{\rm ej} = M_{\rm w} = 0.1M = 1M_{\odot}$, $t_{\rm run} = 10t_{\rm orb} =$ 65 d.  {\bf Top Panel:} Time evolution of luminosities of interest, including: forward shock $L_{\rm f}$, reverse shock $L_{\rm r}$ (dashed black), total shock  $L_{\rm sh} = L_{\rm f} + L_{\rm f}$ (solid black), radiation from shocked disk $L_{\rm w}$ (brown), bolometric $L_{\rm bol}$ (blue), R-band $\nu L_{\nu}$ (red).  {\bf Bottom Panel:} other quantities of relevance, including the shock radius $R_{\rm sh}$ normalized to the semi-major axis (blue), luminosity-weighted temperature of fast ejecta $T_{\rm avg}$ (red), ``photosphere" temperature $T_{\rm ph}$ (black), luminosity-weighted average opacity of ejecta shells $\kappa_{\rm avg}$ (brown)  }
 \label{fig:example}
\end{figure}

Figure~\ref{fig:example} shows an example calculation for binary of mass $M = 10M_{\odot}$, semi-major axis $a = 31R_{\odot}$, orbital period $t_{\rm orb} =$ 6.5 d, escape velocity $v_{\rm esc} = 350$ km s$^{-1}$, slow wind velocity $v_{\rm w} = v_{\rm esc}/4$ and covering fraction $f_{\Omega} = 0.3$, mean dynamical ejecta velocity $\bar{v}_{\rm ej} = v_{\rm esc}$, maximum ejecta velocity $\bar{v}_{\rm max} = 3v_{\rm esc}$, ejecta mass $M_{\rm ej} = M_{\rm w} = 0.1M = 1M_{\odot}$, and pre-dynamical mass-loss timescale $t_{\rm run} = 10t_{\rm orb} =$ 65 d.  

The velocity of the swept-up shell $v_{\rm sh}$ (bottom panel, orange line) decreases over the first few weeks to a value $\lesssim 150$ km s$^{-1}$, midway between that of the fast ejecta and slow wind.  Over the same time interval, the shell/shock radius $R_{\rm sh}$ (bottom panel, blue line) moves outwards to a radius $R_{\rm sh} \gtrsim 10-100a \approx 1-10$ AU (bottom panel, blue line).  The shocks radiate a luminosity (top panel, black line) that peaks at very early times and decreases thereafter as both the shell velocity, and the density of the pre-dynamical outflow ahead of the shock, drops as the shocks propagate outwards.  This radiation leaks out of the equatorial torus into the polar region occupied by the fast ejecta, with a luminosity $L_{\rm w}$ (top panel, brown line) that is slightly delayed from the shock power due to the timescale required for vertical radiative diffusion through the torus (this timescale decreases as the shock propagates outwards to regions of lower density).  

The first peak in the optical light curve $\nu L_{\nu}$ (bottom panel, red line) occurs on a timescale of several days and is powered by the initial thermal energy of the fast layers of the dynamical ejecta, which expand freely along the polar axis of the binary (``cooling envelope'' emission).

However, the second broader peak in the emission at $t \sim 40-60$ days is powered by diffusion of the shock power from greater depths.  The light curve shape is ``bumpy" due to the sensitive dependence of the opacity on the density and temperature at different layers of the ejecta.  The estimated photosphere temperature $T_{\rm ph} \approx T_{\rm eff}$ (bottom panel, black line) decreases from $\gtrsim 6000$ K near the first peak, to $\sim 3000$ K near the second peak.  Starting around day 80, the fastest ejecta shell with $v_{\rm ej} = v_{\rm max}$ reaches a temperature (regulated by irradiation from within) below the value $\approx 1500$ K required for dust nucleation, while the luminosity-averaged temperature of the ejecta reaches this temperature somewhat later, around day 120.  Once dust forms in the ejecta, the bolometric luminosity from shock power can still remain relatively high; however, the effective temperature of the emission will drop considerably due to the formation of a much larger dust photosphere, strongly reducing the optical flux ($\S\ref{sec:dust}$).  

Figure \ref{fig:param} shows a range of models, calculated under different assumptions about the mass and semi-major axis of the binary, as well as the timescale of pre-dynamical mass loss $t_{\rm run}$.  The light curves are seen to produce secondary peaks on timescales ranging from $\approx 30-150$ days with characteristic luminosities $\sim 10^{39}-10^{40}$ erg s$^{-1}$.  In the cases of more massive, extended stars with relatively large values of $t_{\rm run}$, the shape of the light curves - in particular, the presence of a primary and secondary peak - qualitatively resemble the fiducial case.  

In the case of a less massive more compact binary (orange line in Fig.~\ref{fig:param}), the secondary peak is suppressed.  This behavior is due to the strong drop in opacity that occurs around 10 days due to the recombination of the ejecta (as shown by a spike in the emission).  As discussed in the next section, the timescale of the delayed secondary peak present in most models is set by radiative diffusion from the deep inner layers; however, in this case, the shock power is insufficient to keep the ejecta ionized to late times, thus precluding the delayed release of energy responsible for shaping the shock-powered peak.  The details of the light curve (in particular the sudden emission spike) should not be trusted in this case, because our simple model cannot follow the recombination front.  However, we may nevertheless conclude that a prominent shock-powered secondary peak will be suppressed in events where the ejecta recombines prematurely, even if the plateau-shaped light curve preceding it is powered largely by the shocks.\footnote{A similar internally-powered plateau is descrobed by \citet{Sukhbold&Thompson17} in the context of Type IIP supernovae  powered by a central magnetar engine.}  As discussed further in $\S$\ref{sec:discussion}, this provides one explanation for the lack of secondary peaks in some LRN such as V1309 Sco.

\subsection{Approximate Scaling Relations}
\label{sec:scaling}

The basic light curve properties of shock-powered emission in LRN can be estimated using simple analytic arguments, at least in cases where the shock power is sufficient to keep the ejecta ionized at the nominal time of the secondary peak.

We begin by estimating the velocity of the cold central shell.  The radiative shock interaction is momentum-conserving, so the shell velocity equals the ratio of the swept-up momentum $p_{\rm sh}(t) \approx M_{\rm ej}\bar{v}_{\rm ej}f_{\Omega} + M_{\rm w,sh}(t)v_{\rm w}$ and the swept-up mass $M_{\rm sh}(t) \approx f_{\Omega}M_{\rm ej} + M_{\rm w,sh}(t)$,
\be
\frac{v_{\rm sh}}{\bar{v}_{\rm ej}} = \frac{1}{\bar{v}_{\rm ej}}\frac{p_{\rm sh}}{M_{\rm sh}} \approx \frac{1 + (v_{\rm w}/\bar{v}_{\rm ej})(M_{\rm w,sh}/f_{\Omega}M_{\rm ej})}{1 + M_{\rm w,sh}/(f_{\Omega}M_{\rm ej})} ,\,\,\,\, 
\ee
where the swept-up mass of the pre-dynamical wind is given by
\begin{eqnarray}
M_{\rm w,sh}(t)  \approx 
\left\{
\begin{array}{lr}  M_{\rm w}(R_{\rm sh}/v_{\rm w}t_{\rm run}) \approx M_{\rm w}(t/t_{\rm end}),
 & t \lesssim t_{\rm end} \\
M_{\rm w},  &
t \gtrsim t_{\rm end}.\\
\end{array}
\right\}
\label{eq:Mwsh}
\end{eqnarray}
and $t_{\rm end} =  t_{\rm run}(v_{\rm w}/v_{\rm sh})$ is the time required for shock to reach the outer edge of the pre-dynamical outflow.  Although the time evolution of $v_{\rm sh}$ is complicated by the implicit dependence of $M_{\rm w,sh}$ on $v_{\rm sh}$, we find that for fiducial parameters its value varies by less than a factor of two from its asymptotic value of
\be
\frac{v_{\infty}}{\bar{v}_{\rm ej}} = \frac{1 + \frac{v_{\rm w}}{\bar{v}_{\rm ej}}\frac{M_{\rm w}}{f_{\Omega} M_{\rm ej}}}{1 + \frac{M_{\rm w}}{f_{\Omega}M_{\rm ej}}} \approx 0.5-1,
\ee
where in the final line we have taken $M_{\rm w} \lesssim  M_{\rm ej}$, $f_{\Omega} = 0.3$, and $v_{\rm w}/\bar{v}_{\rm ej} = 0.25$.  In our analytic estimates that follow we assume a temporarily constant value of $v_{\rm sh} = \xi \bar{v}_{\rm ej}$, where $\xi \approx 0.5-1$.

Emission escapes on the photon diffusion timescale through the fast ejecta, which at the shock radius $R_{\rm sh} \approx v_{\rm sh}t$ is given by
\be
t_{\rm d,sh} \approx \tau_{\rm ej,sh}\frac{R_{\rm sh}}{c} \approx \left(\int_{R_{\rm sh}} \rho_{\rm ej}\kappa_{\rm ej} dr\right)\frac{R_{\rm sh}}{c} \approx \frac{M_{\rm ej}\kappa_{\rm ej}}{6\pi\xi(\bar{v}_{\rm ej}t)},
\ee 
where we have used equation (\ref{eq:rhoej}) for $\delta_{\rm o} = 5$ and $\delta_{\rm i} = 2$.  

If the opacity remains roughly constant in time, then equating this to the expansion time $t$ gives the timescale for the shock-powered emission to rise to its peak \citep{Arnett82}
\begin{eqnarray}
t_{\rm pk} &\approx& \left(\frac{M_{\rm ej}\kappa}{6\pi \xi c \bar{v}_{\rm ej}}\right)^{1/2} \nonumber \\
&\approx& 27\,{\rm d}\,\,\frac{\kappa_{0.3}^{1/2}}{\xi^{1/2}}\left(\frac{M_{\rm ej}}{0.1M}\right)^{1/2}\left(\frac{\bar{v}_{\rm ej}}{v_{\rm esc}}\right)^{-1/2}\left(\frac{M}{M_{\odot}}\right)^{1/4}\left(\frac{a}{10R_{\odot}}\right)^{1/4} \nonumber \\
&\approx& 27\,{\rm d}\,\,\frac{\kappa_{0.3}^{1/2}}{\xi^{1/2}}\left(\frac{M_{\rm ej}}{0.1M}\right)^{1/2}\left(\frac{\bar{v}_{\rm ej}}{v_{\rm esc}}\right)^{-1/2}\left(\frac{M}{M_{\odot}}\right)^{9/20}\left(\frac{a}{10R_{\rm ms}}\right)^{1/4},
\label{eq:tpeak}
\end{eqnarray}
where $\kappa_{\rm 0.3} \equiv \kappa/(0.3 $cm$^{2}$ g$^{-1}$) is normalized to a typical value (Fig.~\ref{fig:example}).  In the second line we have normalized the ejecta mass to a fixed fraction of the total stellar mass; the fast ejecta velocity to the binary escape speed; and the semi-major axis of the binary to 10 times the main sequence stellar radius $R_{\rm ms} \approx R_{\odot}(M/M_{\odot})^{0.8}$.

\begin{figure}
  \centering
  \includegraphics[width=1.\columnwidth]{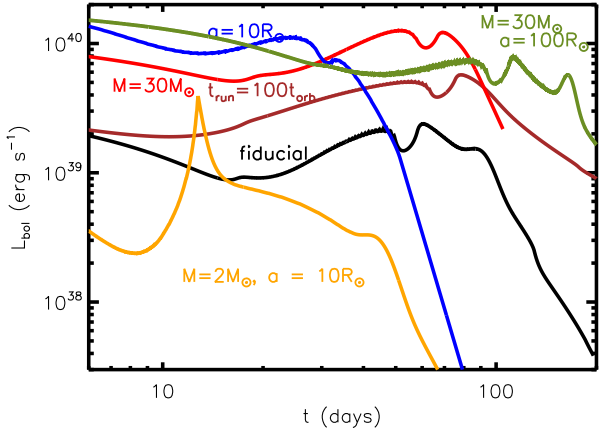}
  \caption{Bolometric light curves for a range of models, showing the common presence of an early-powered light curve peak from the cooling envelope and a later shock-powered secondary maximum.  Models shown include the fiducial case ($M = 10M_{\odot}$, $a = 30R_{\odot}, t_{\rm run} = 10t_{\rm dyn}$; Fig.~\ref{fig:example}), as well as variations with longer pre-dynamical mass runaway time ($t_{\rm run} = 100t_{\rm dyn}$; brown); smaller semi-major axis ($a = 10R_{\odot}$; blue); larger binary mass ($M = 30M_{\odot}$; red); larger binary mass and larger semi-major axis ($M = 30M_{\odot}$, $a = 100R_{\odot}$; green).  Also shown in a less massive star $M = 2 M_{\odot}$ in a more compact binary $a = 10R_{\odot}$ (orange).  This light curve has a qualitatively different shape because the shocks are not powerful enough to keep the ejecta ionized, and the resulting drop in opacity prevents the formation of a second light curve peak (note that our model does not include dust opacity, and so a secondary peak might still form in the infrared).   }
 \label{fig:param}
\end{figure}

All else being equal, equation (\ref{eq:tpeak}) shows that mergers between more massive stars, or evolved stars with radii much larger than their main sequence values, will result in later peaking light curves, consistent with our full numerical solutions (Fig.~\ref{fig:param}). However, overall the dependence on binary parameter is relatively weak, consistent with the factor of $\lesssim 5$ range of observed values $t_{\rm pk} \approx 30-150$ d for $a \sim 2-100R_{\rm ms}$ and $M \sim 1.5-30M_{\odot}$ (Fig.~\ref{fig:LC}, Table \ref{table:1}).  Also note that, by assuming a fixed, relatively high value of the opacity in deriving equation (\ref{eq:tpeak}), we cannot account for cases when the shock heating is insufficient to keep the ejecta ionized (e.g. the 2$M_{\odot}$ model in Fig.~\ref{fig:param}).

If the forward shock dominates the shock power (eq.~\ref{eq:Lf}), then the peak luminosity of the shock-powered second light curve peak is given by (in the limit $v_{\rm sh} \gg v_{\rm w}$)
\begin{eqnarray}
L_{\rm pk} &\approx& L_{\rm f}(t_{\rm pk}) \approx \frac{9}{32 f_{\Omega}}\frac{v_{\rm sh}^{3}}{v_{\rm w}}\frac{M_{\rm w}}{t_{\rm run}}e^{-\left(\frac{v_{\rm sh}t_{\rm pk}}{v_{\rm w}t_{\rm run}}\right)}\nonumber \\
&\approx& 9\times 10^{40}\,{\rm erg\,s^{-1}}\,\,\xi^{3}\left(\frac{f_{\Omega}}{0.3}\right)^{-1}\left(\frac{M_{\rm w}}{0.1M}\right)\left(\frac{v_{\rm w}}{v_{\rm esc}/4}\right)^{-1} \\
&\times& \left(\frac{\bar{v}_{\rm ej}}{v_{\rm esc}}\right)^{3}\left(\frac{t_{\rm run}}{10t_{\rm orb}}\right)^{-1} \left(\frac{M}{M_{\odot}}\right)^{5/2}\left(\frac{a}{10R_{\odot}}\right)^{-5/2}e^{-\left(\frac{v_{\rm sh}t_{\rm pk}}{v_{\rm w}t_{\rm run}}\right)},
\label{eq:Lpeak1}
\end{eqnarray}
where we have used equations (\ref{eq:rhow},\ref{eq:Mdot}).  Normalizing again to the main sequence radius, 
\begin{eqnarray}
L_{\rm pk} &\approx& 9\times 10^{40}\,{\rm erg\,s^{-1}}\,\,\xi^{3}\left(\frac{f_{\Omega}}{0.3}\right)^{-1}\left(\frac{M_{\rm w}}{0.1M}\right)\left(\frac{v_{\rm w}}{v_{\rm esc}/4}\right)^{-1} \nonumber \\
&\times& \left(\frac{\bar{v}_{\rm ej}}{v_{\rm esc}}\right)^{3}\left(\frac{t_{\rm run}}{10t_{\rm orb}}\right)^{-1} 
\left(\frac{M}{M_{\odot}}\right)^{1/2}\left(\frac{a}{10R_{\rm ms}}\right)^{-5/2} e^{-\left(\frac{v_{\rm sh}t_{\rm pk}}{v_{\rm w}t_{\rm run}}\right)},
\label{eq:Lpeak}
\end{eqnarray}
where the argument of the exponential is (assuming $v_{\rm w} = v_{\rm esc}/4$)
\begin{eqnarray}
&&\frac{v_{\rm sh}}{v_{\rm w}}\frac{t_{\rm pk}}{t_{\rm run}} \nonumber \\
&\approx& 3\,\xi^{1/2}\kappa_{0.3}^{1/2}\left(\frac{t_{\rm run}}{10t_{\rm orb}}\right)^{-1}\left(\frac{M_{\rm ej}}{0.1M}\right)^{1/2}\left(\frac{\bar{v}_{\rm ej}}{v_{\rm esc}}\right)^{-1/2}\left(\frac{M}{M_{\odot}}\right)^{-1/4}\left(\frac{a}{10R_{\rm ms}}\right)^{-5/4} \nonumber \\
\label{eq:tpeak2}
\end{eqnarray}

The luminosity of the shock-powered emission increases with the binary mass, and is also generally stronger for collisions between more compact stars or shorter runaway times $t_{\rm run}$, up to the point when the exponential suppression term becomes $\gtrsim 1$.  The most luminous predicted transients are those from high mass stars close to the main sequence $a \approx 2 R_{\rm ms}$ with long pre-dynamical mass-loss timescales $t_{\rm run} \gtrsim 100 t_{\rm orb}$, for which the predicted luminosity is $L_{\rm pk} \gtrsim 3\times 10^{41}$ erg s$^{-1}$, consistent with the most luminous LRN (Fig.~\ref{fig:LC}).  

\begin{figure}
  \centering
  \includegraphics[width=1.\columnwidth]{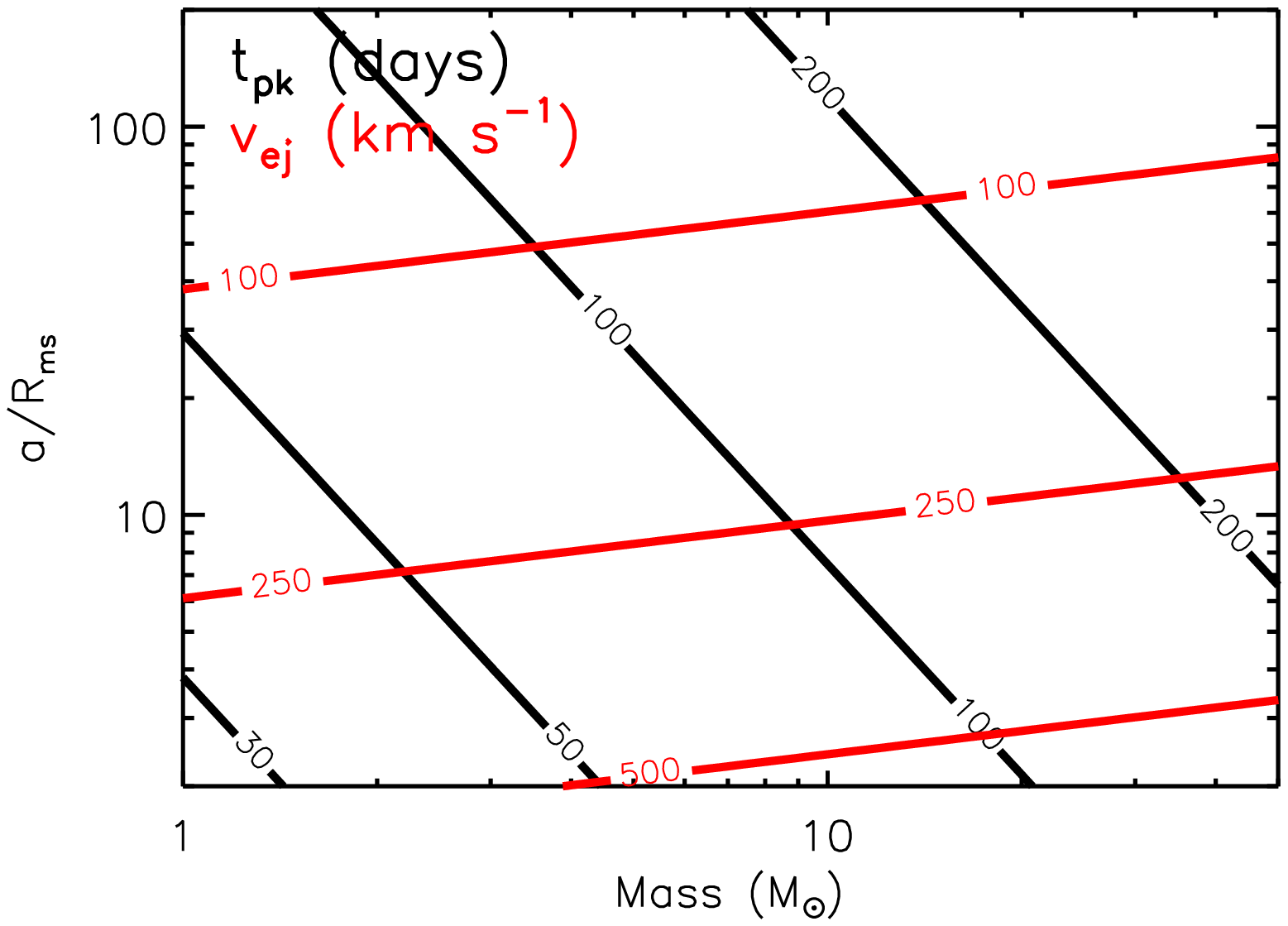}
  \includegraphics[width=1.\columnwidth]{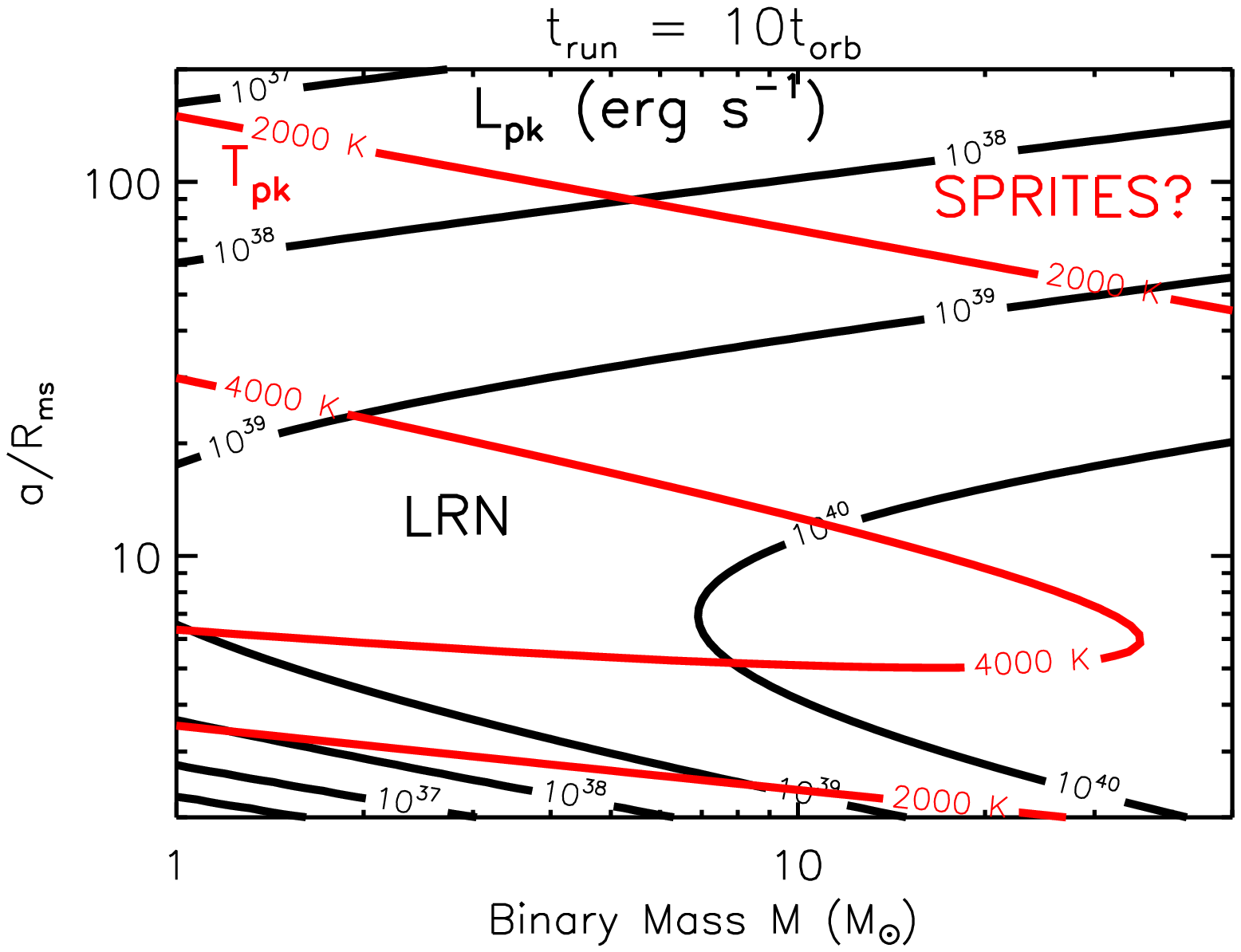}
  \includegraphics[width=1.\columnwidth]{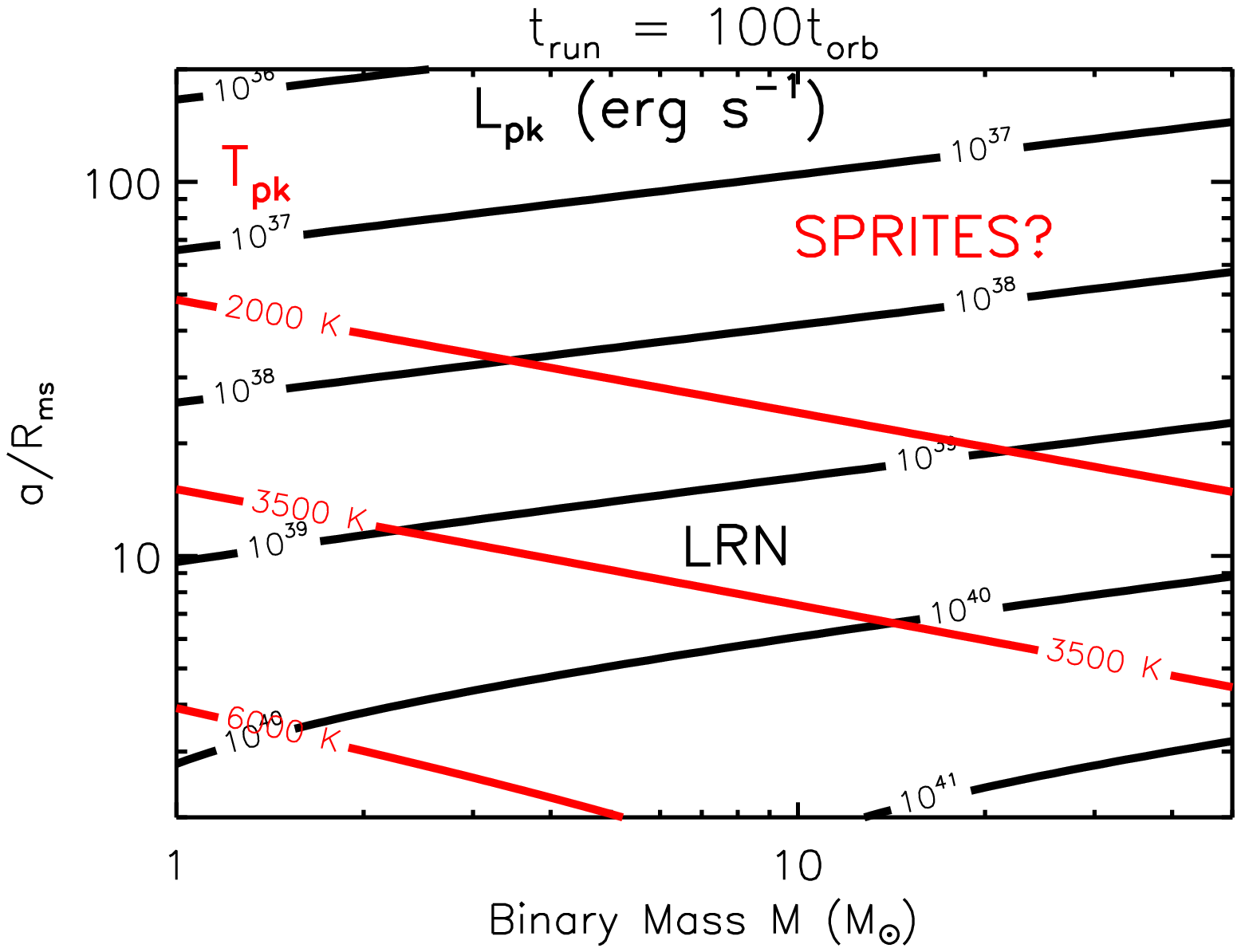}
  \caption{Parameter space of shock-powered emission from stellar mergers, based on the analytic estimates derived in $\S\ref{sec:analytic}$, as a function of binary mass $M$ and semi-major axis $a$; the latter normalized to the main sequence radius $R_{\rm ms}$ of a star of mass $M$.  {\bf Top Panel:} Timescale of peak emission $t_{\rm pk}$ (black contours, eq.~\ref{eq:tpeak}) and mean velocity of fast ejecta $\bar{v}_{\rm ej} = v_{\rm esc}$, calculated for $\xi = 0.5$, $\kappa = 0.3$ cm$^{2}$ g$^{-1}$, $M_{\rm ej} = M_{\rm w} = 0.1M$, $v_{\rm w} = v_{\rm esc}/4$, $f_{\Omega} = 0.3$.  {\bf Bottom Panels:} Peak luminosity $L_{\rm pk}$ (black contours, eq.~\ref{eq:Lpeak}) and estimated ejecta temperature at peak luminosity $T_{\rm pk}$ (red contours), shown for two assumptions about the timescale for pre-dynamical mass loss, $t_{\rm run} = 10t_{\rm orb}, 100 t_{\rm orb}$.  Also marked are the schematic regions of parameter space giving rise to shock-powered emission which peaks at optical wavelengths as luminous red novae (``LRN") versus infrared wavelengths as potential ``SPRITES" (\S\ref{sec:dust}), as delineated by whether the ejecta temperature at the time of the shock-powered peak exceeds that of dust nucleation $T_{\rm pk} \approx 2000$ K.   }
 \label{fig:analytic}
\end{figure}

Figure \ref{fig:analytic} summarizes the above results, showing contours in the parameter space of stellar mass and radius for the peak timescale (top panel) and peak luminosity/temperature (bottom panels), the latter for two assumptions about the timescale of pre-dynamical mass loss, $t_{\rm run} = 10,100t_{\rm orb}$.  

\subsection{Dust Formation}
\label{sec:dust}

 Near- and mid-IR interferometric observations of V838 Mon revealed a resolved flattened disk-like equatorial overdensity of dust with a major axis that aligns with the position angle inferred from polarimetric studies \citep{Chesneau+14}.  Likewise,  V1309 Sco formed a slowly-expanding, dense, and optically-thick dusty envelope during its 2008 outburst \citep{Tylenda&Kaminski16}.  {\it Spitzer} observations of the LRN V4332 Sagittarii \citep{Martini+99} revealed evidence for dust formation \citep{Banerjee+07}.  The observed strengths of emission lines suggest that only a small fraction of the emission from the central object is observed, presumably due to a dusty disk viewed edge on \citep{Kaminski+10}, consistent with evidence from linear polarization for dust scattering in an equatorially-focused disk \citep{Kaminski&Tylenda11}.  
 \citet{Mauerhan+17} also provide evidence, based on the mid-infrared through UV observations, for aspherical dust distribution in SN Hunt248.

Dust formation thus appears to be ubiquitous in LRN at late times, and its inferred geometry may support the shock interaction picture developed in this paper.  The radiative shocks that occur in the equatorial plane of the binary provide a natural location for the observed dust formation.  Gas cools behind a radiative shock from the immediate post-shock temperature of $\approx 10^{5}-10^{6}$ K to $\lesssim 10^{4}$ K, resulting in compression by a factor $\gtrsim 10-100$ from the pre-shock density.  Additional radial expansion is required for the shocked matter to cool further to the temperatures $\lesssim 2000$ K required for dust nucleation.  However, residual concentration of a dense dust-forming region in the binary equatorial plane appears to be a natural prediction of our model.  A similar model for dust formation in equatorially-concentrated radiative shocks was recently proposed for classical novae \citep{Derdzinski+17}.   

Across much of the stellar merger parameter space, the shocks are sufficiently luminous to keep the ejecta hot enough to delay dust formation until after the time of the secondary peak.  However, the bottom panel of Figure \ref{fig:analytic} shows low peak temperatures $T_{\rm pk} \lesssim 2000$K for compact stars in the case of short runaway times $t \lesssim 10t_{\rm orb}$, or from more radially extended stars $a \gtrsim 10R_{\rm ms}$ in the case of long runaway times $t \gtrsim 100 t_{\rm orb}$ (upper right corner of lowest panel).  Mergers with these properties could form a large quantities of dust {\it even prior to the secondary shock-powered light curve}, resulting in the formation of a much larger cooler photosphere due to the high dust opacity and a luminous {\it infrared} transient.  

This subset of mergers may be particularly relevant to the class of infrared transients recently discovered by the ongoing SPIRITs (SPitzer InfraRed Intensive Transients Survey) of nearby galaxies with {\it Spitzer} \citep{Kasliwal+17}, some of which show no detectable optical counterparts and absolute magnitudes at 4.5$\mu$m from $M_{V} \sim -11$ to $-14$.  These so-called SPRITES (eSPecially Red Intermediate-luminosity Transient Events) are characterized by a wide range of photometric evolution rates, corresponding to timescales as short as months or as long as several years.  Our results suggest that stellar mergers of giant stars, with particularly long pre-dynamical mass loss phases ($t_{\rm run} \gtrsim 10-100t_{\rm orb}$), represent an attractive explanation for the SPRITE phenomenon.  These events may have been optically-bright briefly near the point of the dynamical merger, but their time interval as unobscured optical transients could be very short compared to the bulk of the latter shock-powered, infrared-bright phase.  Potentially supporting picture is the discovery of shock-excited molecular hydrogen emission from SPIRITS 14ajc, which \citet{Kasliwal+17} propose could have resulted from the merger of a proto-stellar binary. The object OGLE2002-BLG-360 \citep{Tylenda+13} might represent a low-mass counterpart to SPRITES.

\section{Discussion and conclusion}
\label{sec:discussion}

Previous work has interpreted the late phase of LRN emission as being powered by hydrogen recombination \citep{Ivanova+13b,Macleod+17}.  However, the required ejecta masses in some cases are quite large: \citet{Macleod+17} estimate that the $100-200$ day emission of M101 OT2015-1 requires an ejecta mass of $1-10M_{\odot}$, close to the entire mass of the binary.  In our shock-powered scenario, the same luminosity can be powered by a much lower ejecta mass $\lesssim M_{\odot}$ since the energy per gram of shock-powered emission $\sim v_{\rm esc}^{2}$ greatly exceeds the $\sim 1$ Rydberg per particle supplied by hydrogen recombination.  

If the optical LRN can be explained by the ejection of relatively small amount of mass $M_{\rm ej} \sim 0.1M$ then this would imply that most of the current sample are likely to be bona fide stellar mergers, in which most of the envelope is retained. Furthermore, the progenitors of LRN are consistent with being relatively compact stars near the main sequence and not red (super)giants. Mergers of the latter, more evolved stars are likely to appear significantly redder, dustier, and longer in duration, consistent with the observed properties of SPRITEs \citep{Kasliwal+17}.  Given the limited information available on the final outcome of LRN, it is unclear whether these events completely eject their envelope, leaving behind a tight binary pair. Direct detection of the surviving binary  - or even a single merger remnant - is challenging due to lingering presence (and perhaps continual production) of dust. The prolonged 3.6 and 4.5$\,\mu$m emission from M101 OT2015-1 and NGC 4490-OT \citep{Blagorodnova+17,Smith+16}  hints that a continuum of transient properties may exist between SPRITES and LRN. 

In order to reproduce the light curves of LRN, we require that the total amount of mass loss from the binary during the dynamical and pre-dynamical phases are roughly comparable, i.e.~$M_{\rm ej} \sim M_{\rm w} \sim 0.1M$.  Such a substantial amount of pre-dynamical mass loss is potentially surprising within the current ``common envelope'' paradigm, which dictates that the most important phases of the merger process occurs only once the binary loses co-rotation with - and begins to feel hydrodynamical drag on - the envelope.  Indeed, the pre-dynamical mass loss required by our models would alone be sufficient to carry away the angular momentum required to merge the binary, if this matter leaves the binary with the same specific angular momentum of the L2 point.  

Our results also have implications for the ability of numerical simulation to address the stellar merger problem.  The accumulation of slowly-expanding or quasi-stationary circumbinary wind material, released over many orbital periods, would be very challenging to study using typical common envelope hydrodynamical codes due to long timescales involved, (potentially) insufficient resolution near the stellar surface, and possible radiative/thermal processes that control the runaway of the binary (which are not typically modeled).  Sustained pre-dynamical mass loss furthermore implies that initial conditions for the dynamical phase are likely to be substantially different than currently assumed, due to changes in the envelope of the stars caused by the preceding phase of gradual mass loss.  All of these aspects informing the appropriate ``initial conditions" for numerical studies can be constrained by developing more sophisticated models for shock-powered emission in LRN.  These will more tightly constrain $M_{\rm ej}$, $M_{\rm w}$, and their relative velocity from the timing and luminosity of the secondary light curve peak, combined with constraints on the ejecta velocity structure from spectroscopy.

\citet{Macleod+17} interpret the first peak of M31 LRN 2015 as being due to cooling envelope emission of $\sim 10^{-2}M_{\odot}$ of fast ejecta driven at the onset, and they interpret the later plateau-shaped emission as being powered by hydrogen recombination of a larger ejecta mass $\sim 0.3M_{\odot}$ to power the later recombination-driven plateau.  In our picture, both components of emission result from a singular mass ejection event created during the dynamical phase of the merger.  The low inferred mass of the ``fast" ejecta, which contributes to the earliest emission phase, is a simple byproduct of the typical density profile of an dynamical explosion (eq.~\ref{eq:rhoej}), which places only a small fraction of the total mass at high velocities (as commonly inferred also in Type IIP supernovae). Supporting our interpretation of this event is pre-outburst photometry showing evidence for pre-dynamical mass loss \citep{Dong+15}. Ultimately, the luminosity and duration of the first light curve peak constrains the initial thermal energy and quantity of the fastestest expanding dynamically-ejected matter.
 
We have discussed scenarios for producing the second light curve peaks, which are commonly but not ubiquitously observed in LRN (Fig.~\ref{fig:LC}).  LRN without second peaks include V1309 Sco and M31 LRN 2015.  In $\S\ref{sec:results}$ we described how a second plateau could possibly be avoided if the shocks are insufficient to keep the ejecta hot enough to stay ionized until the nominal Arnett peak timescale (eq.~\ref{eq:tpeak}).  In this case the resulting drop in opacity results in rapid cooling of the ejecta and dust formation, precluding a true optical peak (though the plateau-shaped emission prior to this point may still be shock-powered, and a secondary peak may still occur in the infrared).  Another possible explanation for lack of a plateau is due to the influence of viewing angle.  V1309 Sco was likely observed near the binary equatorial plane (\citealt{Tylenda+11}), in which case equatorially-concentrated dust formation in the pre-dynamical mass loss or shock-compressed gas ($\S\ref{sec:dust}$) could result in obscuration of the shock-powered emission peak (which diffuses out through the lower density polar regions; Fig.~\ref{fig:cartoon}).  In the latter case the dense shell would need to reside near or above the optical photosphere of the fast ejecta; future work should address possible emission line signatures of these exposed shocks. 

From the sample of LRN light curves in Figure \ref{fig:LC}, an apparent correlation exists between the shock-powered emission peak and the time interval between the peaks.  Comparing equation (\ref{eq:tpeak}) and (\ref{eq:Lpeak1}), for shock-powered emission we predict $t_{\rm pk} \propto M^{1/4}a^{1/4}$ and $L_{\rm pk} \propto M^{5/2}a^{-5/2}$.  A positive correlation between $t_{\rm pk}$ and $L_{\rm pk}$ would therefore suggest that binary mass instead of semi-major axis is a primary candidate for the variable driving this relationship.  However, this apparent correlation could also be shaped by selection effects: mergers of wide (large $a$) binaries may suffer early dust formation, making them observable only as SPRITES instead of optically-luminous LRN (Fig.~\ref{fig:analytic}).  

Early observations of some LRN indicate that their colors prior the first light curve maximum are redder than near the peak \citep{Kurtenkov+15,Kankare+15}. This is unusual because explosions typically evolve from blue to red, and our model is no exception (Fig.~\ref{fig:example}). We speculate that this early red emission could be another signature of extended pre-dynamical equatorial mass loss not included in our model.  Either one is directly observing the pre-dynamical wind at early times (which is naturally cooler than the recently shocked dynamical ejecta), or the outer regions of the pre-dynamical wind are absorbing and reprocessing emission from the dynamical ejecta before being overtaken.  We note that similar red-to-blue-to-red light curve evolution is observed also in the SN IIn 2011ht \citep{Mauerhan+13}, lending credence to our explanation for LRN since pre-explosion mass loss is the hallmark feature of supernovae IIn. 

Finally, although we focused on a simple mass loss history prior to the dynamical phase of the merger, in principle a more complex series of mass ejection events could occur if the common envelope evolution occurs in several stages (e.g.~\citealt{Ivanova+13b}).  In such cases, more than two broad light curve peaks are possible, similar to the complex light curve structure observed in some Type IIn SNe.  The bimodal picture of a fast dynamical explosion plowing into a slow steady wind adopted in this work is also almost certainly too simplified; the transition between these two phases will in reality be more gradual than we have assumed.

\section*{Acknowledgements} 

We thank Lars Bildsten, Mansi Kasliwal, and Morgan MacLeod for valuable discussions.  BDM gratefully acknowledges support from the National Science Foundation (AST-1410950, AST-1615084), NASA through the Astrophysics Theory Program (NNX16AB30G) and the Fermi Guest Investigator Program (NNX15AU77G, NNX16AR73G), the Research Corporation for Science Advancement Scialog Program (RCSA 23810), and the Alfred P.~Sloan Foundation.

\bibliography{stellarmergers}

\label{lastpage}

\end{document}